\documentclass[sigconf, natbib=false]{acmart}

\AtBeginDocument{%
  }

\setcopyright{none}

\RequirePackage[
  datamodel=acmdatamodel,
  style=acmnumeric,
  ]{biblatex}

\usepackage{comment}
\usepackage{subcaption}
\usepackage{fancyvrb}

\begin{document}

\title{Objcache: An Elastic Filesystem over External Persistent Storage for Container Clusters}

\author{Takeshi Yoshimura}
\affiliation{\institution{IBM Research - Tokyo}\city{Tokyo}\country{Japan}}

\author{Tatsuhiro Chiba}
\affiliation{\institution{IBM Research - Tokyo}\city{Tokyo}\country{Japan}}

\author{Sunyanan Choochotkaew}
\affiliation{\institution{IBM Research - Tokyo}\city{Tokyo}\country{Japan}}

\author{Seetharami Seelam}
\affiliation{\institution{IBM T. J. Watson Research Center}\city{Yorktown Heights, NY.}\country{USA}}

\author{Hui-fang Wen}
\affiliation{\institution{IBM T. J. Watson Research Center}\city{Yorktown Heights, NY.}\country{USA}}

\author{Jonas Pfefferle}
\affiliation{\institution{IBM T. J. Watson Research Center}\city{Yorktown Heights, NY.}\country{USA}}

\begin{abstract}
Container virtualization enables emerging AI workloads such as model serving, highly parallelized training, machine learning pipelines, and so on, to be easily scaled on demand on the elastic cloud infrastructure. 
Particularly, AI workloads require persistent storage to store data such as training inputs, models, and checkpoints.
An external storage system like cloud object storage is a common choice because of its elasticity and scalability.
To mitigate access latency to external storage, caching at a local filesystem is an essential technique.
However, building local caches on scaling clusters must cope with explosive disk usage, redundant networking, and unexpected failures.
We propose objcache, an elastic filesystem over external storage.
Objcache introduces an internal transaction protocol over Raft logging to enable atomic updates of distributed persistent states with consistent hashing.
The proposed transaction protocol can also manage inode dirtiness by maintaining the consistency between the local cache and external storage.
Objcache supports scaling down to zero by automatically evicting dirty files to external storage.
Our evaluation reports that objcache speeded up model serving startup by 98.9\% compared to direct copies via S3 interfaces.
Scaling up with dirty files completed from 2 to 14 seconds with 1024 dirty files.
\end{abstract}

%



\settopmatter{printacmref=false} 
\renewcommand\footnotetextcopyrightpermission[1]{} 
\pagestyle{plain}

\maketitle

\section{Introduction}

Container orchestration systems (e.g., Kubernetes) are an essential cloud infrastructure to manage a wide variety of workloads in a computing cluster.
The infrastructure automatically manages the lifecycle of each software deployment in a cluster and thereby enables better resource allocation, crash recovery, and cluster scaling for cost saving~\cite{Borg-EuroSys20}.
It optimizes the utilization of general computing resources such as CPUs and memory but also high-performance devices such as GPGPUs, 100-Gb networks, and NVMe.
As a result, we can easily build an infrastructure for emerging workloads, e.g., distributed machine learning for training and inference of large language models with pipelines.

However, the current container infrastructures face challenges in persistent state management.
Foremost is that they increase the complexity of infrastructure management by having to decouple storage domains and perform data backups at node maintenance.
One way to address this is to use external storage such as cloud object storage or fully managed data stores.
However, external storage tends to have limitations imposed by cloud vendors with non-POSIX interfaces such as S3 API.
Also, the decoupling of the data and computing domains increases the end-to-end service latency for frequent reads on hot data and intermediate data writes shared among distributed jobs.

As a result, end-user applications often have a common pattern of local cache at the local filesystem (FS) to read and write external storage.
It downloads files to the local FS, passes them to the core logic of the application, generates outputs, and uploads them later.
This pattern is also observed in multiple frameworks for different application domains such as the genome workflow engine~\cite{Cromwell} and model serving~\cite{Triton}.
However, implementing this simple pattern is more complex than expected since the local cache needs to be efficient, consistent, and fault-tolerant.

The existing wrapper FSs for external storage can be a local cache for such use cases, but unfortunately, none of them satisfies the requirements of container clusters.
Existing adapter tools such as Lustre hierarchical storage management (HSM)~\cite{Lustre} and Spectrum Scale's active file management (AFM)~\cite{SpectrumScale-AFM} enable applications to transparently synchronize files between an existing distributed FS and external storage.
However, it is unclear whether they can be directly deployed to container clusters with cluster scaling, since they are still coupled to local storage to manage persistent data.
Agni~\cite{Agni-SoCC19} is another candidate for this use case, but it also depends on shared persistent logs for consistency management.
S3FS~\cite{S3FS} and Goofys~\cite{Goofys} can run write-though cache for S3-like storage with local FS.
The elasticity of Snowflake storage~\cite{Snowflake-NSDI20} implies that its write-through cache can easily adapt to cluster scaling, but it is still inefficient for distributed pipelines to write data shared among nodes.
LPCC~\cite{LPCC-SC19} utilizes local storage for cache by leveraging HSM, but it does not allow partial updates of files because of its data synchronization with tools outside of detailed FS internals.

In this work, we propose objcache, an elastic FS over external storage for containers.
As a distributed FS, objcache maintains the consistency among local storage in a cluster and also between local and external storage such as COS buckets.
From different perspectives, we build hierarchical storage over local storage and COS buckets so that users can consistently read/write objects on COS as POSIX files.
Objcache maps objects stored at COS buckets (e.g., s3://bucket/key) to files (e.g., /opt/bucket/key) and caches and shards them at the local storage in a cluster.
We prototype objcache as FUSE and CSI drivers with an operator for OpenShift clusters.

A key enabler of the consistency among local and external storage is our transaction protocol.
We utilize Raft log~\cite{Raft-ATC14} to enable transaction restarts after a crash by log replay.
The protocol is similar to the existing one over Paxos~\cite{lamport2001paxos} (e.g., MegaStore~\cite{Megastore-CIDR11} and Spanner~\cite{Spanner-OSDI12}) but we add external storage as transaction participants.
Objcache precisely tracks updates of each chunk of all the objects at local storage and safely commits changes after confirming the completion of remote uploads.
Consequently, the uploaded files can be reconstructed later by re-downloading them from external storage.
A cluster that runs objcache as a storage domain can quickly delete or add a part or the entirety of it even as it maintains a persistent state.
The protocol also ensures the isolation and ordering of racy writes to follow the POSIX consistency semantics~\cite{POSIX-write}.
Note that we do not currently enable replication, but we are planning to support replications for high availability in the future version of objcache.

The contributions of this paper are as follows.
\begin{itemize}
\item \textbf{Transaction protocol for distributed FS states over external durable storage}:
Files are partitioned and distributed as metadata and multiple chunk objects.
Objcache keeps the consistency of concurrent reads/writes on those objects with a two-phase commit protocol over Raft.
It maintains two levels of cache in cluster-local storage and node-local memory.
Our experiment showed that the cache tiering speeded up sequential reads by 193\% to 1115\% compared to S3FS.
\item \textbf{Distributed filesystem that can scale down to zero}: Objcache supports zero scaling by synchronizing all dirty files to external storage.
This is because we can fetch non-dirty files from external storage again.
Zero scaling can simplify cluster maintenance.
At scaling events, objcache drops non-dirty files to reduce the amount of data migration between local nodes to restore only dirty files.
As a result, our experiment showed that scaling up with dirty files completed from 2 to 14 seconds per node addition with 1024 dirty files.
Scaling down with dirty files required from 2 seconds to 6.8 seconds per node removal.
\item \textbf{Shared filesystem over COS}:
As a specific but important use case, objcache also can replace too common and redundant software components to copy data between local storage and COS.
Compared to existing approaches such as S3FS and Goofys, objcache eliminates duplicated file contents in a cluster with sharding and consistent hashing while ensuring their strict consistency.
Write-back cache at local storage optimizes object uploads/downloads with inter-node pipelines.
Users can utilize both the high speed of local storage and the huge capacity of COS.
Our experiment on a training workload showed that objcache improved the performance of model loading by 24\% and checkpointing by 274\% compared to S3FS.
\item \textbf{Performance observation of two consistency models in distributed FS}:
Objcache supports read-after-write and close-to-open consistency models.
We examined the performance difference in our objcache prototype.
Close-to-open showed better throughput for overall workloads but the simplicity of client code for read-after-write improved the performance of random read workloads.
\end{itemize}

This paper opens with research challenges on existing work (Section~\ref{sec: motivation}).
Then, we describe the high-level architecture of objcache and how to use it in OpenShift (Section~\ref{sec: usage}).
We then present the cache design (Section~\ref{sec: design}) and implementation details (Section~\ref{sec: impl}), followed by our evaluation of objcache (Section~\ref{sec: exp}).
We conclude with a brief summary and mention of future directions (Section~\ref{sec: conclusion}).

\section{Related work}
\label{sec: motivation}

Objcache has three key aspects: shared POSIX FS over COS, distributed storage tiering, and zero/quick scaling.
This section describes each aspect with motivating use cases along with the research challenges of existing approaches.

\subsection{Shared POSIX FS over COS}
\label{sec: motivation: posix}

COS is now a common data lake, but there are many container and serverless workloads that still work with POSIX file interfaces.
For example, a genome workflow engine~\cite{Cromwell} can run on container platforms such as AWS Batch with high parallelism,
but it still copies inputs to local FS since many of the analysis tools depend on local files.
A popular distributed machine learning framework~\cite{Pytorch} provides APIs for model loading and checkpointing with local FS so that users can easily try their training on a shared FS.
A model serving infrastructure for inference~\cite{Triton} also maintains model storage in local FS to build a local cache for object storage.

The primary reasons for copying files to local FS are the requirements of local cache, limited support of new data sources due to legacy applications, and/or convenience for general users.
We predict this trend will not change because of the lack of portability of local FS and external storage like S3.
Especially for local caching of COS, developers need to deeply understand the behavior of distributed storage to correctly handle different consistency models among local and external storage.

Existing POSIX FSs over COS~\cite{S3FS} \cite{Goofys} \cite{Agni-SoCC19} can mostly mitigate the above difficulties, but all of them still assume weak consistency (close-to-open or NFS semantics) between local storage and/or COS.
Users need to be aware of weak consistency (read-after-write) and confirm their distributed jobs do not require strict consistency.
An example use case of the strict consistency is a compilation job for fusing CUDA kernels that assumes shared FS to use files for inter-process communication.
If users need to run such code with strict consistency in a container cluster,
they need to set up shared FS for it and copy files manually or use adapter tools~\cite{Lustre} \cite{SpectrumScale-AFM}.

The efficiency of metadata management in distributed FS tends to become an issue because of the difficulty in managing their consistency~\cite{Metadata-Survey-TPDS22}.
A popular design to ensure metadata consistency is to build dedicated metadata servers like HDFS~\cite{HDFS-MSST10}, GFS~\cite{GFS-SOSP03}, and CephFS~\cite{CephFS-OSDI06}.
However, centralized metadata servers implicitly rely on a weak consistency model to directly update objects at data servers.
We enforce strict consistency with an internal transaction protocol across cluster nodes with a decentralized architecture.
HopsFS~\cite{HopsFS-FAST17} also leverages NewSQL and its transaction for encapsulating FS operations for metadata management at scale.
Wang et. al.,\cite{CFS-EuoroSys23} points out and mitigates scalability issues of transaction processing between metadata services and data stores.
Objcache integrates a transaction processing system to its FS internal to unify multiple storage systems in a cluster.
By unifying storage systems, we can reduce distributed locking by merging multiple operations if they run in the same node.
Currently, objcache does not optimize metadata management with external information including performance profiles, but it can be promising for future performance improvement as shown in CFS~\cite{CFS-SIGMOD19}.

Objcache supports the consistency models of read-after-write and close-to-open while supporting partial, random writes for both, as explored in past research~\cite{CAPFS-FAST05}.
If necessary, users can choose the strict POSIX semantics so that concurrent processes can read/write the same file with expected ordering.
This means that objcache enables IPCs with files as if processes in a cluster were in the same physical node.
This property lets developers easily run complex distributed jobs while also using simple Shell/Python scripts for waiting for events in a cluster with common FS operations instead of socket programming.

\subsection{Distributed Storage Tiering}
\label{sec: motivation: storagetiering}

Data sizes keep increasing rapidly.
Distributed machine learning consumes multi-modal files including natural language texts, images, sounds, and videos on the GB scale and beyond.
Generated files can be Python pickle files and structured data formats such as Parquet and Apache Arrow.
Many existing works for distributed storage tiering focus on optimal interfaces for key-value stores~\cite{Snowflake-NSDI20} \cite{Pocket-OSDI18} \cite{Cachew-ATC22}.
Objcache builds distributed storage tiering while preserving the strict semantics of POSIX file interfaces.

As discussed in Section~\ref{sec: motivation: posix}, many real-world applications build a local cache to improve performance and synchronize it with external durable storage such as S3 buckets.
However, when building a local cache, it is difficult to avoid exceptional errors.
The local cache must be aware of the capacity of local storage and enforce cache replacement at high disk pressure.
Multiple users may not want to share the same storage area for isolation but nevertheless are forced to share a single physical storage.
Local cache may also need to be fault tolerant to handle node crashes.
In a high-level view, objcache composes hierarchical storage with local and external storage so that users can easily aggregate local cache while handling complex issues around cache management.
For example, objcache aggregates the capacity of local storage in a cluster by partitioning and distributing a large file as multiple chunks and metadata.

Production FSs already offer adapters to automate the synchronization of local and remote storage.
Lustre HSM~\cite{Lustre} and SpectrumScale AFM~\cite{SpectrumScale-AFM} can run copy tools to transparently access remote storage.
LPCC~\cite{LPCC-SC19} leverages Lustre HSM to manage the consistency of local and external storage.
However, those copy tools are plugin components and cannot fully manage the internal states of existing distributed FSs.
We design a distributed FS from scratch with the awareness of storage tiering.
As a result, objcache introduces an internal transaction protocol to enable atomic updates of partial files and dynamic scaling.

Another approach is to let container schedulers optimize access patterns to remote storage and local cache for application efficiency.
The effectiveness of scheduler awareness of local cache is shown by locality-aware scheduling in existing distributed computing frameworks~\cite{MapReduce-OSDI04} \cite{Spark-HotCloud10} \cite{Ray-OSDI18}.
Stateful serverless infrastructures~\cite{Pocket-OSDI18} \cite{Cloudburst-VLDB20} \cite{Boki-SOSP21} also provide storage and scheduler integration.
SiloD~\cite{SiloD-EuroSys23} optimizes container scheduling for better caching of remote storage leveraging workload characteristics of deep learning.
Objcache focuses on its FS design but can be extended to work with schedulers for adaptive prefetching and improve scalability.


\subsection{Quick/Zero Scaling}
\label{sec: motivation: scaling}

Storage makes cluster management difficult.
Unlike other components in a container cluster, storage needs to maintain persistent states tightly coupled with physical storage.
Existing FSs already support scaling, but they still assume keeping data somewhere in local storage.
This property is not suitable for the immutability of container clusters and it raises challenges in maintenance and cluster scaling for cost saving.
Also, machine learning users may start from minimum nodes for interactively testing and inspecting raw data, but they may finally run workflows at scale.

To the best of our knowledge, objcache is the first distributed FS that supports quick/zero scaling.
LambdaObjects~\cite{LambdaObjects-HotStorage22} enables elastic storage that can be co-located to computing nodes, but it relies on its object interface for serverless.
Faa\$T~\cite{FaaST-SoCC21} enables elastic cache with autoscaling, but it provides shared memory APIs instead of FS interfaces.
Stateful serverless infrastructures~\cite{Pocket-OSDI18} \cite{Cloudburst-VLDB20} \cite{Boki-SOSP21} \cite{FaaST-SoCC21} also design their architecture with the awareness of cluster scaling,
but they provide get/put-like interfaces.
FS interfaces require more complex data management than simple key-value stores since FS supports directory structures, partial writes/reads, renaming, and so on.
Distributed FSs have challenges in their elasticity since they often need to partition and distribute internal states across nodes.
Objcache solves the challenge by reusing its internal transaction protocol for atomic updates of distributed states.

Another approach of FS scaling is to synchronize files to external storage as discussed in Section~\ref{sec: motivation: storagetiering}.
Lustre HSM~\cite{Lustre} and SpectrumScale AFM~\cite{SpectrumScale-AFM} can automate synchronization.
However, they reuse existing distributed FSs to build local cache, and thus, the elasticity of underlying FS becomes a problem in this case.
In particular, scaling events requires data migration in underlying FS states,
but it is not clear if the adapter tools can efficiently work with internal states.
Objcache integrates the FS and the synchronization logic into a single software system to efficiently handle data migration at scaling events.

\section{Architecture of Objcache}
\label{sec: usage}

\begin{figure}[t]
  \centering
  \begin{subfigure}{0.48\linewidth}
    \centering
    \includegraphics[width=\linewidth]{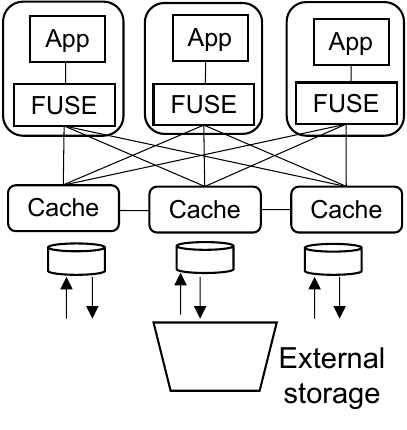}
    \caption{Detached deployment}
    \label{fig: detached_architecture}
  \end{subfigure}
  \hfill
  \begin{subfigure}{0.48\linewidth}
    \centering
    \includegraphics[width=\linewidth]{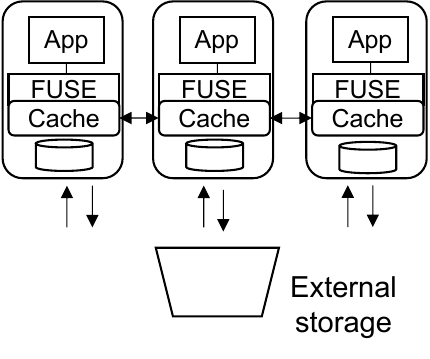}
    \caption{Embedded deployment}
    \label{fig: embedded_architecture}
  \end{subfigure}
  \caption{Architecture overview.}
  \label{fig: architecture}
\end{figure}

Our prototype of objcache runs as processes of FUSE (Filesystem in Userspace) and cache servers distributed in a cluster (Figure~\ref{fig: detached_architecture}).
We also support an embedded deployment of FUSE and cache (Figure~\ref{fig: embedded_architecture}) to improve performance although it has limited management flexibility we discuss later.
For the detached deployment, users can start a FUSE process to create a mount point in a node and let the user's applications request file operations such as \verb|read| and \verb|write| from the FUSE process to cache server ones.
Cache servers coordinate with each other to expose external storage as a single FS in a cluster.
We do not limit the number of groups of FUSE and cache servers.
If a user needs to isolate accesses to different external storage, different objcache groups can run in the same cluster.
This section describes how users can construct, use, manage, and scale their cache in a cluster before diving into the internal details described in later sections.

\subsection{Deployment in a Container Cluster}

Objcache regards popular container orchestration systems, Kubernetes and OpenShift, as its primary platform so that it can reuse many cluster management functionalities.
Users can configure every objcache process (i.e., FUSE and cache servers) in a cluster manually, but we also provide an automation tool with the Kubernetes operator pattern to reduce the burden on the users.
It utilizes the Kubernetes container storage interface (CSI) to expose itself as a Storage Class in a Kubernetes cluster.
Kubernetes enables users to easily mount it in their containers.
Consequently, we need to run CSI controller and node processes in addition to FUSE and cache servers in a Kubernetes cluster.
As a common situation, we assume a Kubernetes cluster with two user roles: administrators and general users.

For detached deployment in Figure~\ref{fig: detached_architecture}, an administrator deploys the objcache operator with two custom resource definitions, \verb|ObjcacheCsiDriver| and \verb|ObjcacheCluster|.
These custom resources direct how the operator deploys, manages, and configures the entire objcache and CSI processes in a Kubernetes cluster.

\begin{figure}[t]
  \centering
  \includegraphics[width=0.9\linewidth]{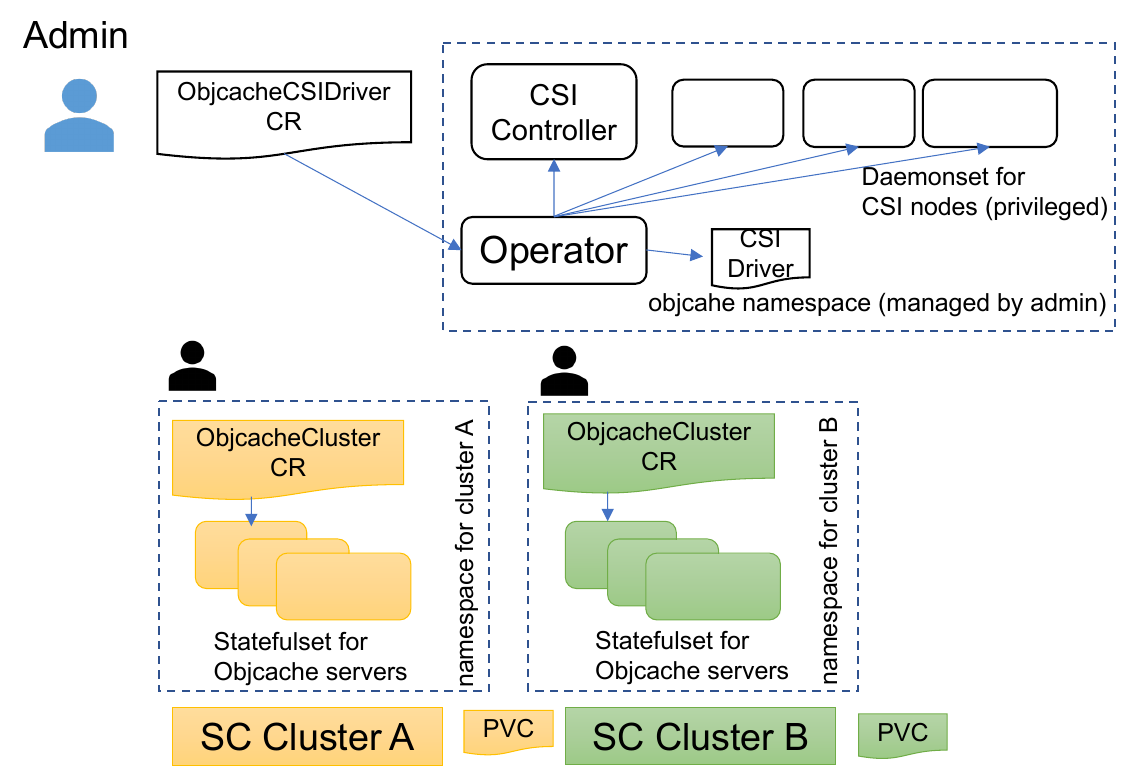}
  \caption{\textbf{Deployment overview}.}
  \label{fig: deployment}
\end{figure}

Figure~\ref{fig: deployment} shows how objcache clusters are deployed in a Kubernetes cluster.
First, an administrator deploys a custom resource of \verb|ObjcacheCsiDriver| to configure the objcache CSI driver in a cluster.
It contains a service account to grant access rights for the required resources in a Kubernetes cluster.
The administrator can also specify the node affinity to deploy objcache CSI elements if they need to limit and isolate a part of the cluster nodes to deploy.
The operator starts deployment of the objcache CSI controller and a daemon set of the objcache CSI node.
The CSI controller manages the information of the storage provision and attachment for user containers in a cluster.
Kubernetes automatically watches the controller deployment and restarts/reschedules it in the event of crashing or node removal.
The daemon set of the CSI node automatically starts at every node in a cluster to launch FUSE in local nodes in the event of attachment requests from containers.
The request contains the information for the address of the attached objcache servers deployed by users later.
The deployment of \verb|ObjcacheCsiDriver| and the service account needs administrative privilege since Kubernetes CSI elements require privilege to access the host FS to control mounts for containers.

\begin{figure}[t]
  \centering
  \begin{subfigure}{0.48\linewidth}
    \centering
    \includegraphics[width=\linewidth]{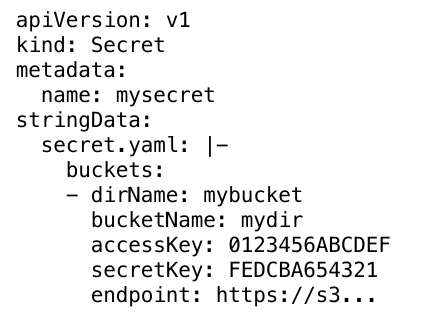}
    \caption{Secret for a S3 bucket.}
    \label{fig: yaml-secret}
  \end{subfigure}
  \hfill
  \begin{subfigure}{0.48\linewidth}
    \centering
    \includegraphics[width=\linewidth]{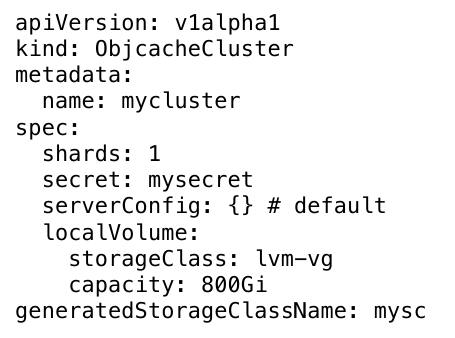}
    \caption{Cluster configuration.}
    \label{fig: yaml-cluster}
  \end{subfigure}
  \caption{Example configurations. Users configure an objcache cluster with YAML files.}
  \label{fig: yaml}
\end{figure}

General users can deploy an \verb|ObjcacheCluster| custom resource to let the operator start, manage, and scale their cache servers with detailed configurations including their node affinity.
In the configuration, users must specify the credential information for external storage as a Secret resource in a Kubernetes cluster (Figure~\ref{fig: yaml-secret}).
The example secret contains a YAML file that specifies containers to mount a COS bucket s3://mybucket at (mount directory)/mybucket with an S3 credential.
Users can add multiple buckets at different directories.
The credential information is applicable to any S3-compatible storage systems including AWS Simple Storage Service, Google Storage, IBM Cloud Object Storage, and other third-party software such as MinIO and Ceph.

Figure~\ref{fig: yaml-cluster} shows a simple configuration with a cache server.
A user can dynamically scale up/down the cache server by changing the number of shards.
The operator automatically starts or shuts down cache servers as multiple Kubernetes Stateful sets with the number of configured shards.
The configuration specifies a storage class to provision 800-GB logical volume in a cluster node where the cache server is deployed.
The operator finally generates a new storage class for the objcache cluster.
Briefly speaking, in Kubernetes clusters, users can create persistent volume claims pointing to the storage class to let containers mount the storage class.
The mount request is finally propagated to the CSI nodes that an administrator deployed and it launches FUSE processes to connect general applications and cache servers that the user deployed.

Embedded deployment provides a single custom resource that merges \verb|ObjcacheCluster| and \verb|ObjcacheCsiDriver|.
It deploys an embedded image with FUSE, CSI, and cache as a single container.
Both deployments offer the same functionality we present in this paper, but performance characteristics are different due to eliminated network communication within the same node.
Embedded deployment has operational limitations compared to detached deployments.
An objcache cluster creation requires the administrative privilege of a cluster.
A crash of the embedded FUSE requires pod restarts since FUSE processes cannot keep mount connections to the Linux kernel after their terminations.

\subsection{Basic File Accesses}
\label{sec: fileaccess}

Applications can transparently access files on external storage through objcache FUSE and cache servers.
For path lookup, we follow the approach of existing wrapper FSs for S3-like storage such as S3FS~\cite{S3FS} and Goofys~\cite{Goofys} although we allow mounting multiple buckets at a single FUSE instance.
Objcache directly maps a key in external storage to a file in a mounted FS.
It regards a character '\verb|/|' as a path delimiter and regards a key with the suffix \verb|/| as a directory.
For example, a key \verb|a/b/c/d.txt| in a COS bucket \verb|bucketA| is mapped to a file at \verb|(mount directory)/bucketA/a/b/c/d.txt|.
Cache servers at first maintain only the root directory \verb|/| with directories representing specified external storage.
If a user requests a key with an FS operation, e.g., \verb|read| and \verb|write|, the FS recursively retrieves child directories and files in the target external storage from the top directory to the leaf file before starting the requested operations on it.

\subsection{Consistency Model}
\label{sec: consistency}

Objcache needs to synchronize two different groups of ``remote'' storage:
specifically, the storage in a cluster and the external storage outside of it (e.g., COS buckets).
The cluster-local storage is distributed across nodes, so we also need to maintain the consistency between cluster-local storage.

\textbf{Consistency of External Storage}: Cache servers maintain a write-back cache for external storage and dirty objects, which are objects with updated contents that have not been uploaded yet.
Therefore, compared to no cache or write-through cache, we relatively relax the consistency of files at the external storage and allow the potential loss of updated contents at specific failures.
Note that our design minimizes the occurrence of data losses and enforces the failure atomicity as described in Section~\ref{sec: transaction}.

Currently, objcache does not automatically check if the current cache is outdated from the version of the external storage.
Therefore, an objcache cluster may read stale contents when the target external storage is updated by other means such that a user directly uploads files with the S3 API and more than two isolated objcache clusters update the same file in the same bucket.

\textbf{Consistency of Files}: Network FSs like objcache need to maintain consistency among distributed storage in a cluster.
POSIX semantics require read-after-write consistency, in which an FS must reflect a \verb|write()| if a \verb|read()| occurs after the \verb|write()| \cite{POSIX-write}.
To avoid stale reads due to remote updates, every \verb|write()| must not be buffered locally and must immediately reflect the latest FS state.
Every \verb|read()| must check the latest FS state if there are any updates at that moment.
As a result, this semantics operation sacrifices I/O performance, although it is often too strict for many parallel jobs to follow.

Close-to-open consistency relaxes the requirements of reflection timing, where a \verb|read()| can return stale values at the time of \verb|open()| and the reflection of \verb|write()| can be delayed until \verb|close()|.
Many parallel distributed jobs can run with close-to-open consistency, but some workloads may need to check file updates.

Objcache is configurable for both read-after-write and close-to-open consistency models for local cluster storage.
Every cache server tracks the latest content of its local cache with transactions.
In contrast, FUSE processes cannot easily check inodes updated by remote FUSE processes in a cluster since they may do buffering at their local memory or the Linux page cache.
Therefore, objcache changes its consistency model by enabling or disabling buffering and the Linux page cache for FUSE at a local node.
After starting a transaction to update FS states, it ensures the ordering of FS operations as required by the POSIX semantics.
POSIX semantics also define the concurrent behavior of other FS operations than \verb|read| and \verb|write|  (e.g., \verb|mkdir|) but we follow the strict ordering constraints with our transaction protocol.

\subsection{Failure Recovery}

Objcache internally runs Raft-based logging for supporting log replays after a crash.
Crashes can occur at any failures in hardware and software, 
but objcache attempts to recover from them with the capability of Raft logging.
Objcache may block the operations at a crash but can recover from the failure after the failed instance restarts and replays the local log.
We do not currently enable replications, but we are also planning to support it in the future version of objcache for high availability.
When a failure or timeout occurs in the middle of updating FS states, all the participants are rolled back to the healthy states before the transaction starts to retry the same request from the beginning.

Currently, objcache cannot resume its execution from the cases where mismatched checksums at disk contents are detected.
In this case, new transactions cannot proceed anymore and all the servers need to be restarted. 
This means that we roll back to the time of the last upload to COS.

\section{Cache Design}
\label{sec: design}

\begin{figure}[t]
  \centering
  \includegraphics[width=\linewidth]{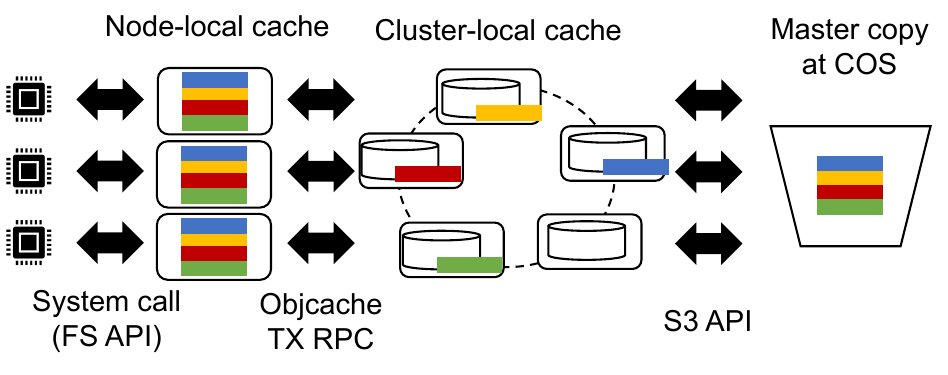}
  \caption{\textbf{Cache layers and internal APIs}. Objcache fills an interface gap between local and external storage with additional layers. Objects in external storage are partitioned into fix-sized chunks and distributed with consistent hashing.}
  \label{fig: layers}
\end{figure}

Figure~\ref{fig: layers} shows the high-level cache architecture of objcache.
The objcache FS manages two levels of cache, node-local and cluster-local cache to manipulate system calls for file operations (e.g., \verb|open|, \verb|read|, and \verb|close|) on objects in external storage.
The node-local cache is an in-memory data store to ensure the consistency described in Section~\ref{sec: consistency} and also to minimize the latency of reading cluster-local objects that are recently accessed locally.
The cluster-local cache maintains a file as fix-sized chunks (e.g., 16 MB) and file metadata (e.g., dirty, timestamp, and file size) to distribute them across persistent storage at cluster nodes with consistent hashing.
The cluster-local cache uploads itself as files in an external storage at the event of synchronization such as \verb|fsync| calls and expiration of dirty objects.
Persistent storage at local nodes utilizes Raft-based logging for write-ahead logging to be replayed after crash events.

\subsection{Distributed Metadata and Data Chunks}

Cluster-local cache manages on-disk/in-memory inodes for every inode with single metadata and multiple data chunks.
Objcache distributes chunks to cluster-local servers as done by HDFS~\cite{HDFS-MSST10} and GFS~\cite{GFS-SOSP03}.
However, we also distribute metadata to cluster-local servers to build an elastic, flat architecture and maintain a node-local cache.

The on-disk inode manages an inode ID, inode size, dirtiness, inode type, file permissions, the last modified timestamp, and so on.
The in-memory inode contains a transaction lock and mapping of an inode and physical key at external storage (e.g., bucket key for cloud object storage).
The mapping can reduce name lookups for parent directories when uploading/downloading files at external storage.
A node-local cache exposes an FS to applications by maintaining the mapping of inodes in the local operating system and a cluster-local cache.

The data part of an inode is partitioned into multiple chunks with fixed sizes (e.g., 16MB).
The benefit of data partitioning is that it can easily utilize the storage capacity and network bandwidth of multiple nodes.
Users can write/read larger data than the capacity of cache storage by streaming data fragments from/to external storage.
Objcache can utilize parallel uploads/downloads for synchronizing files among local and external storage.
The chunk also keeps track of its file offset, length, and outstanding writes with target offset and length.
For overwrites, outstanding writes may contain a special entry with a key at external storage.

Currently, we simplify directory management with metadata by regarding them as special files with child inodes and names.
They are also distributed across cluster-local caches, and thus, every name lookup potentially accesses remote nodes.

Distributed chunks and metadata pose two challenges: 
1) how to select a responsible node for metadata and chunks to store those objects in local storage
and 2) how to ensure the atomicity/consistency of racy read/writes on the same offset and/or beyond chunk boundaries.
Also, a common challenge when building efficient distributed POSIX FS is the requirement to ensure the atomicity of operations on different nodes:
specifically, filling multiple chunks, updating object size, and adding the file to a directory.
We solve the former with consistent hashing (Section~\ref{sec: consistenthashing}) and the latter with an internal transaction protocol (Section~\ref{sec: transaction}).

\subsection{Consistent Hashing}
\label{sec: consistenthashing}

Consistent hashing is a common technique to manage consistent metadata for distributed FSs~\cite{Metadata-Survey-TPDS22}.
The basic idea of consistent hashing is to utilize the fact that a hash function given the same key returns the same value.
It calculates the hash values for nodes (\verb|H(N1)| for node \verb|N1|) to associate a key to node \verb|N1| if the hash value of the key is within \verb|[H(N1), H(Na)]|, where \verb|H(Na)| is the minimum hash value for a cluster node other than \verb|N1| (i.e., \verb|N1|'s \textit{successor}).
If \verb|H(N1)| was the maximum value among cluster nodes, it associates a key within \verb|[H(N1), infinite]| and \verb|[-infinite, H(Nb)]|, given that \verb|H(Nb)| is the minimum hash value among cluster nodes.

Objcache utilizes consistent hashing to determine the responsible node for an inode by using an inode ID as a key for the hash function.
The hash value of an inode ID is also used for determining the responsible node for the first chunk (i.e., offset 0) of the inode.
The node for the other chunks is calculated using the combined value of an inode ID and the target chunk offset.
For example, we concatenate two integers with a special delimiter \verb|/| to generate a key as a string.

Consistent hashing enables us to simplify the cluster architecture since we do not need to track every location of metadata and chunks.
Objcache does not have a primary-worker architecture like many other distributed FSs but runs as a peer-to-peer architecture.
Every node-local cache can eliminate redundant node lookups to implement inode operations.
In contrast, cache servers need to synchronize the node list of a cluster every time they handle a cluster scaling, 
which is a relatively rare event compared to inode operations.

Every cluster-local cache maintains and synchronizes the list of all the running participant processes to calculate consistent hashing.
A node list is updated only if cluster scaling occurs.
Cluster scaling can be divided into two operations: a join and a leave of a participant node.
Both operations require atomic updates of a node list in a cluster although it is distributed across nodes.
We reuse consistent hashing and our transaction protocol to synchronize the node list.

\subsection{Cluster Reconfiguration and Data Migration}
\label{sec: clusterreconf}

\begin{figure}[t]
  \centering
  \includegraphics[width=\linewidth]{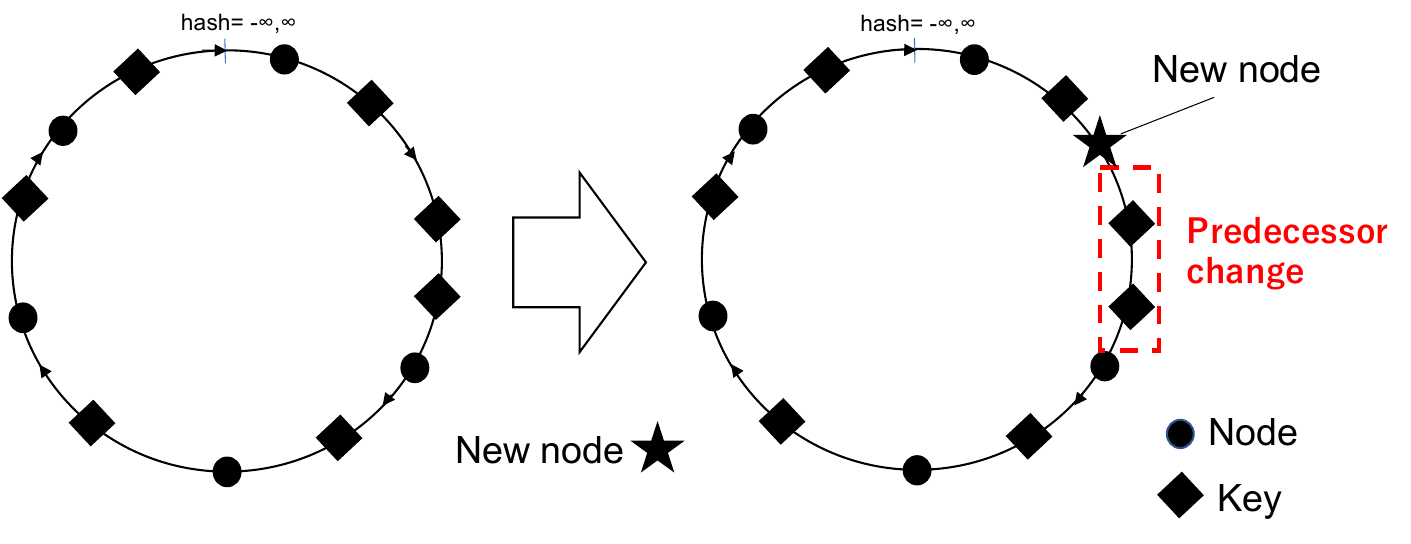}
  \caption{Data migration}
  \label{fig:rehashing}
\end{figure}

A participant joining requires that an operator specifies the endpoint address for another participant unless it is the very first one in a cluster.
Therefore, the operator remembers and passes the address for a running participant (this requirement is straightforward for Kubernetes cluster setups).
The second and later joiners request updating the node list but also migrating chunks and metadata.
Note that the property of consistent hashing minimizes migration at cluster scaling since a node join/leave affects only its successor/predecessor node in consistent hashing.
If a node \verb|N4| joins to cluster nodes \verb|N1|, \verb|N2|, and \verb|N3|, given \verb|H(N1) < H(N2) < H(N4) < H(N3)|, cluster-local cache needs to migrate a part of the metadata and chunks at \verb|[H(N2), H(N3)]| to \verb|N3|.
In other words, metadata and chunks at \verb|[-infinite, H(N2)]| and \verb|[H(N3), infinite])| are not migrated (Figure~\ref{fig:rehashing}).

Also, we can avoid migrating non-dirty objects since external storage already keeps the latest copy of these.
When a node leaves, it waits until all the proceeding updates complete and makes FS read-only.
Then, it initiates uploads of dirty chunks and shutdowns without migration.
Likewise, when a node requests a join to a participant, the participant asks every node to check if the joined node changes the responsible metadata and chunks of the node.
If so, the node makes FS read-only and migrates objects to the joiner.
After migration completes, the node makes itself writable again and processes pending requests (node-local cache retries).
Note that objcache needs to track the dirtiness of not only metadata but also chunks.
A chunk can be assigned to a node that is leaving or changes the predecessor after a node join.

At a request for a node addition or removal, objcache starts a transaction at a node selected by consistent hashing for a special key.
It can process every reconfiguration request one by one although it may cause redundant transmissions of dirty metadata and chunks.

Nevertheless, objcache preserves directory structures during cluster reconfiguration even if we track the dirtiness of directories.
When the metadata for a new file is migrated to another node, we still need to recognize it under the directory at a cluster-local cache.
A problem is that a parent directory of the directory (i.e., a grandparent) for a new file may not be modified.
In that case, objcache may overwrite the grandparent directory and it finally wipes out updated contents with the old version of a parent directory that does not have the new file.
Therefore, a transaction for node join and leave must ensure migrating cluster-local cache for directory metadata if their predecessor changes.

Node-local cache lazily maintains the node list copied from a cluster-local node at the time of an FS operation.
We ensure every FS request from a node-local cache has the version of the node list so that a cluster-local cache can validate it if they use the same node list.
When cluster-local cache reports its inconsistency, node-local cache pulls the latest one and retries its request.
Therefore, a node-local cache can immediately terminate and notify unmount events to applications only if they operate nothing on the FS.
As a result, an external operator (human operator or automated cluster orchestration such as a Kubernetes operator) can easily start cluster scaling, which starts new cache processes or terminates running ones.

\subsection{Internal Transaction Protocol}
\label{sec: transaction}

Objcache internally has a transaction protocol that enforces a simple two-phase commit~\cite{Transaction-Book} to update metadata and chunks in atomic.
For describing our transaction protocol and server roles, we use the same terminologies as common distributed transaction protocols: client, coordinator, and participant.
A \textbf{client} is a server thread that runs within a FUSE instance and requests a coordinator for inode operations.
A \textbf{coordinator} is a server thread that enforces the atomic updates of metadata and chunks by requesting a series of RPCs to \textit{participants}.
A \textbf{Participant} is a server that prepares, commits, or aborts operations at its persistent logs.

This protocol ensures atomic updates of any objects at more than two different nodes.
Therefore, all the file operations become atomic even if they require updating multiple files and directories, which makes POSIX semantics complex.
In other words, we do not use this protocol for updates at a single node.
We utilize this transaction protocol to implement POSIX file operations (Section~\ref{sec: impl}),
and also we reuse it for updating the node list in a cluster (Section~\ref{sec: clusterreconf}).

When FUSE receives an application's request for a file operation (e.g., open, write, read), the transaction client creates a transaction request.
Then, it determines predecessor nodes for updated metadata and chunks with consistent hashing as described in Seciton~\ref{sec: consistenthashing}.
It sends the request to the node for metadata as a transaction coordinator (and may retry others if a timeout occurs).

The two-phase commit (2PC) protocol separates an update request into ``prepare'' and ``commit'' requests.
A coordinator sends prepare requests to all the participants,
and these participants then acquire the locks for updated states and save the updated version in the redo log without updating in-memory working states.
If the coordinator confirms that all the participants have returned a success, it sends a commit request to all the participants.
They append the commit record to their persistent log with in-memory working states updated and unlocked.

The coordinator can abort at an error and may retry its operation later.
The coordinator requests an abort to all the participants before a commit and reports it to the client.
Each participant appends an abort record and unlocks the states without modifying them.
The client propagates persistent errors such as \verb|ENOENT|, which racy file operations may cause.
For transient errors such as timeouts and known temporal outages, the client retries the request.

This transaction protocol ensures the ordering of racy updates even if they try to update the area crossing a chunk boundary and/or the same offset.
For example, if a client updates an inode with chunks Ca1 and Ca2, and another client updates the same inode with chunks Cb1 and Cb2 at the same offset as the former client, 
then readers should observe the inode with either Ca1-Ca2 or Cb1-Cb2.
The pair of preparation and abort means ensuring the same order of the update executions at all the predecessor nodes can be ensured, although we cannot guarantee the order of request arrivals.

General 2PC is known to have a weakness in the commit phase:
namely, if nodes failed during a commit, the transaction cannot proceed further.
In objcache, Raft logs (Section~\ref{sec: raftlog}) can handle such situations, since the node can be consistently restarted with replayed logs or else delegated to followers.
After a log replay, objcache can resume committing or aborting updates with write-ahead log records.

\subsection{Transaction ID}

Objcache needs to handle the possibility of request duplication.
In objcache, clients independently maintain timeout for a coordinator in case of its silent failures,
so a client may try to restart a live but very slow coordinator, which potentially causes deadlocks due to duplicated requests.
This can occur when heavy requests cause I/O stalls.

We extend sequence numbers for the original Raft to handle duplicated requests~\cite{Raft-Dissertation}.
Every transaction request has a unique ID with a tuple of client ID, sequence number (SeqNum), and transaction sequence number (TxSeqNum).
The client ID is the unique ID of the transaction client within a FUSE instance.
The SeqNum is a monotonic, local clock at the client.
The TxSeqNum is a fixed sequence number that each coordinator maintains so that a coordinator can restart a series of RPCs with the same IDs to make itself idempotent.

A client executes a coordinator RPC in accordance with a user's file operation.
It passes the ClientId and SeqNum and a coordinator generates TxSeqNum for its operations locally or remotely.
For usual operations, we do not have to care about this, but things get quite complex if errors or timeouts happen.
With TxId, the client can restart the coordinator and participants with the exactly same TxId for the same operation.
As a result, participants can determine duplicated requests with the passed TxId.
When objcache detects a duplicated request, it replies with old results as done in the Raft RPCs.

\subsection{Raft Logs}
\label{sec: raftlog}

Objcache fault tolerance depends on the capability of the Raft logging mechanisms~\cite{Raft-ATC14} to store transaction records.
The original Raft ensures the consistency of entries appended to logs~\cite{Raft-ATC14}, but it does not define its entries itself.
We define its entry structure so that it can be used with our transaction processing (Section~\ref{sec: transaction}) and various inode operations.
As a result, our transactions and FS states are implemented as a series of state machines in Raft logs.

\begin{figure}[t]
  \centering
  \includegraphics[width=\linewidth]{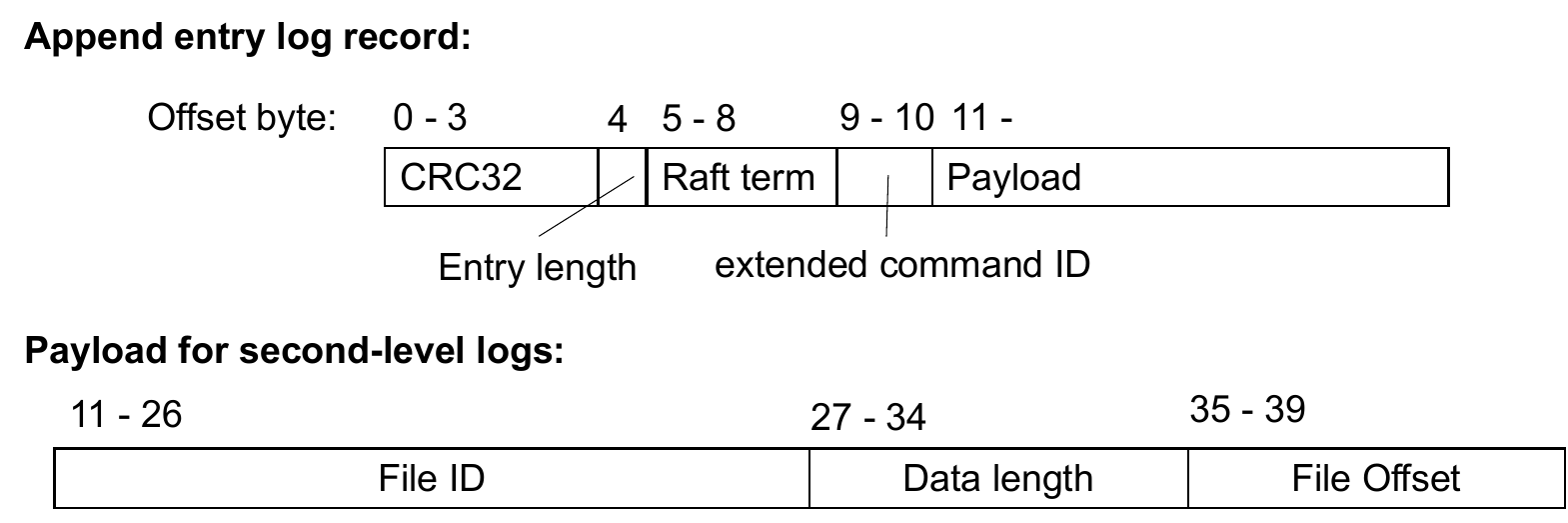}
  \caption{\textbf{Raft log format and an example payload for second-level logs}.}
  \label{fig: raftlog}
\end{figure}

Figure~\ref{fig: raftlog} shows an example format of extensible appended entries to safely preserve transaction and command information.
The figure is still abstracted to support various inode operations, but it already contains control information for transaction processing.
It contains a leader's term, transaction command type ID, checksum, length, and extended payload.
The extended payloads contain pre-defined formats according to the command ID.
For example, a command to commit metadata update has the transaction ID to let a transaction manager start committing an update and unlocks it.
We implemented 72 variants of state machine commands in total.

Figure~\ref{fig: raftlog} also shows a payload for second-level logs.
Objcache extends log entries so that they have pointers to second-level logs to easily support varied-sized log entries.
Currently, we implement logging at files at local FS, and thus, the pointer for a log entry consists of the file ID, offset, and length.
For example, a file write is directly appended to a predecessor's second-level log.
The primary log records a tuple of file ID, offset, and length.
When an application writes new offsets in a file, cluster-local caches append log entries for updated states related to metadata and/or chunks.
Outstanding writes are appended to predecessors' logs with a file ID, offset, and length as a log entry.
A transaction for flushing writes updates a copy of the current metadata to update the file size and append a log entry for it.
The predecessors for the written chunks copy current log entries for chunks and update them with the pointer of the second-level log entries for outstanding writes.

When the Raft protocol elects a follower to be a leader, it resumes incomplete transactions.
Therefore, a client/coordinator can eventually complete its requests by retrying one of the servers in a group even if there are server failures.
If a coordinator/participant detects itself as a follower at the beginning of its RPC, it rejects a request with a likely leader's server ID.

\section{Implementation}
\label{sec: impl}

We prototyped objcache with around 30,000 lines of Go (version 1.19).
Specifically, we extended a FUSE library to support low-level notifier requests to invalidate local cache and post-processing callbacks after each read operation (around 100 lines of Go).
We confirmed it runs on OpenShift 4.11.9 with Red Hat Enterprise Linux CoreOS 4.11 (Linux kernel 4.18.0).
We also directly tested FUSE on Ubuntu 22.04.2 (Linux 5.15.0).

\begin{figure}[t]
  \centering
  \includegraphics[width=\linewidth]{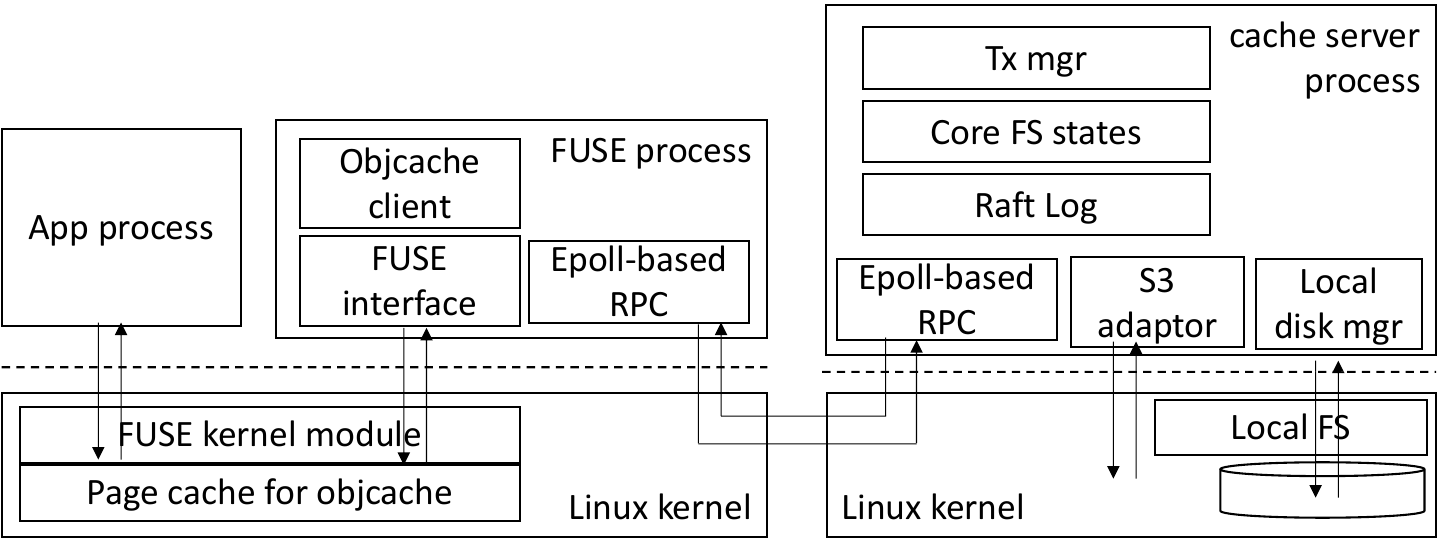}
  \caption{\textbf{Internal components}.}
  \label{fig: impl_architecture}
\end{figure}

An overview of our implementation of node-local and cluster-local caches is shown in Figure~\ref{fig: impl_architecture}.
A FUSE process works with the Linux page cache if the weak consistency (close-to-open) is enabled.
Each process communicates with the other using Epoll-based remote procedure calls (RPCs).
The S3 adapter downloads/uploads files in COS buckets.
Transaction manager utilizes RPCs to interact with participants to accomplish requested file operations.
FS states are tracked in Raft logs on local FS (e.g., XFS) for fault tolerance.
We ensure using direct I/O and \verb|fsync()| after every log appends, but our prototype relies on the durability of local FS.

\subsection{open()}

At \verb|open()|, FUSE recursively looks up a series of metadata on cluster-local storage from the root to the target leaf inode.
If cluster-local storage failed to find metadata in its local storage, it then retrieves a key in external storage by joining the parent path and the name of its inode.

\subsection{fsync()}

Objcache uploads dirty inodes to S3 buckets when \verb|fsync()| is called or configured flush intervals are passed.
Every cache server tracks successor inodes in consistent hashing and starts a persisting transaction for expired dirty files.
At \verb|fsync()|, FUSE processes also start the persisting transaction.
The persisting transaction runs with a 3S bucket as a participant in addition to predecessor nodes for metadata and chunks.
S3-like storage exposes multipart upload (MPU) APIs with begin, add, abort, and commit to upload a key in parallel.
Those APIs may fail for any reason, and thus, we need to execute them before the commit phase.

\begin{figure}[t]
  \centering
  \includegraphics[width=\linewidth]{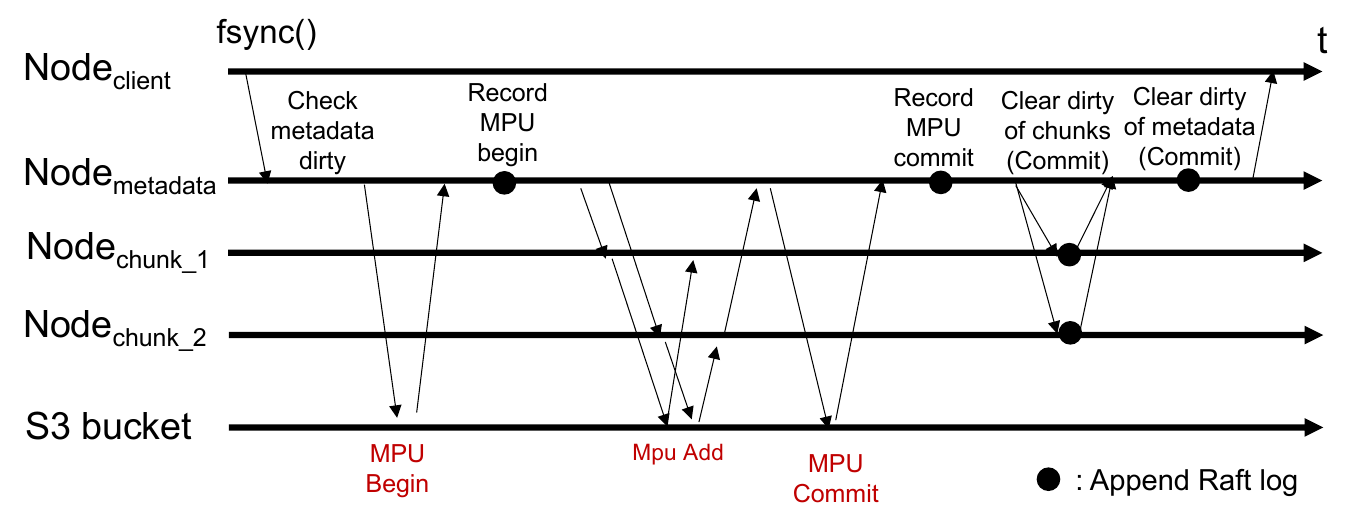}
  \caption{Mixed transaction with S3 multipart uploads (MPU). Objcache records Raft logs in case of failures at black dots in the figure.}
  \label{fig: 2pc_fsync}
\end{figure}

We describe the transaction of \verb|fsync()| in Figure~\ref{fig: 2pc_fsync}.
First, the objcache client in a FUSE process determines a metadata predecessor with consistent hashing with the target inode.
It calls an RPC to launch a coordinator for persisting metadata.
The coordinator checks the dirtiness of the metadata and calls the MPU Begin to create an upload key.
The key is recorded at the Raft log before the MPU Commiit since we need to abort the upload in case of a failure.
The coordinator then calls RPCs to let chunk predecessors upload their chunks to the S3 bucket with the MPU Add.
After the upload is finished, the coordinator calls the MPU Commit and records the uploaded inode at Raft logs.
During the commit phase, it calls RPCs to clear dirty flags to participants for chunks, and then it clears the flag for metadata.
Note that a failure between the MPU Commit and recording the log may result in uploading the same content twice because of the log replay.

As an optimization, we use the PutObject API instead of MPU for a small inode with less size than a single chunk (e.g., 16MB).
We regard the predecessor for the first chunk of an inode to be the same as the metadata by passing the same key for those calculations (i.e., passing the inode number for the hash function).
In that case, we can skip recording the MPU begin, and also the commit record can be a single Raft log write since the coordinator is a single participant in the transaction.

\subsection{write()}

\verb|Write()| is the most interesting operation in terms of consistency management.
We execute a transaction to guarantee atomic updates of metadata and chunks even if multiple chunks are updated at a time.
A naive implementation may first take a pessimistic lock for metadata, transfer written buffers, and commit with the inode size updated.
It can ensure the ordering of racy writes by locking the metadata.
However, the metadata lock can be scalability issues due to the metadata lock during slow buffer transfers.

We implement \verb|write()| as more than two separate transactions.
In particular, updating chunks is treated as a special transaction.
At write(), a client directly transfers chunk updates to a participant without updating metadata.
Then, the client can start a flush transaction to update chunks and the new size of metadata.
Chunk updates are separately recorded as outstanding writes and they are committed to being a new version of the chunk.
As a result, we can shorten a critical section to update an inode.
When the close-to-open consistency model is enabled, an objcache client can update the inode with a single transaction with multiple \verb|write()|.
For small inodes, we can optimize a transaction as a single log append like the optimization for \verb|fsync()|.
Client failures may lose the information of outstanding writes, but this does not break the consistency of any files (it may lead to the creation of an orphan chunk, which a recovery procedure like fsck should reclaim).

Objcache supports random overwrites even though the internal Raft log is append-only.
The outstanding buffers for \verb|write()| are recorded as a partial write with a target inode, offset, length, and written data.
If writes skip some offsets in a file, it adds a special outstanding write to the chunk with the key for external storage before other \verb|write()|.
The logic for \verb|read()| downloads the file fragment and merges it with the written data.

\subsection{Other POSIX File Operations}

Objcache provides transparent access to files at both local cache and external storage.
If the file is not cached at cluster nodes, each predecessor node downloads its range of an inode that corresponds to their chunks from external storage.
Clients can utilize network bandwidth for multiple nodes to download a single file.
If a client requests a file path with the suffix '/', objcache retrieves a file list under it.
COS does not directly support directory structures, but it allows specifying a prefix to filter out keys that are not under a path.

Truncating and deleting a file set a deleted flag to chunks and metadata with zero size and a dirty flag.
Then, the procedure of \verb|fsync| calls delete commands to external storage.
Delete flags enable us to avoid racy read or write fails to proceed operations in atomic.

\subsection{Cluster Reconfiguration}
\label{sec: impl_clusterreconf}

We implemented cache servers to work with cluster autoscaling of computing nodes.
Therefore, the join logic needs to minimize potential reads after scaling up.
On the other hand, the leaving logic can take time but ensure the consistency of local and external storage.
As a result, transactions for scaling up migrate dirty metadata, chunks, and directories that change their predecessor.
On the other hand, scaling down uploads dirty metadata and chunks that change their predecessor, but directories are still transferred to their predecessor.

\section{Experiments}
\label{sec: exp}

In this section, we report the quantitative performance results of objcache.
Our experiments examined the baseline performance of read and write, scalability to multiple nodes, and elasticity under dynamic cluster scaling.
We measured the performance with synthetic FIO workloads and more realistic workloads with a model serving and fine-tuning of large language models.
All the experimental results in this section were an average of more than ten runs of the same experiment.
All the experiments ran either a 4-node OpenShift cluster with A100 GPUs or a 36-node cluster with Ubuntu virtual servers in IBM Cloud.

In the OpenShift cluster, each node had eight 80GB A100 GPUs, two 2nd Generation Intel Xeon Scalable processors (20 cores of Cascade Lake with Hyperthreading), 1.5TB of DRAM, four 3.2TB NVMe drives, and two 100G network interfaces.
As external storage, we used a regional bucket (US East) from IBM Cloud Object Storage.
Each experiment section describes more details of the resources allocated to each job.

Ubuntu virtual servers in IBM Cloud were \verb|mx2d-4x32| instances,
which had 4 virtual CPUs of Cascade Lake, 32 GB RAM, 8Gbps networks, and a 150-GB solid state drive.
It ran in a Tokyo region, and thus, external storage for this cluster was a regional bucket in a Tokyo availability zone.

\subsection{Cache Tiering}

\begin{figure}[t]
  \centering
  \includegraphics[width=0.6\linewidth]{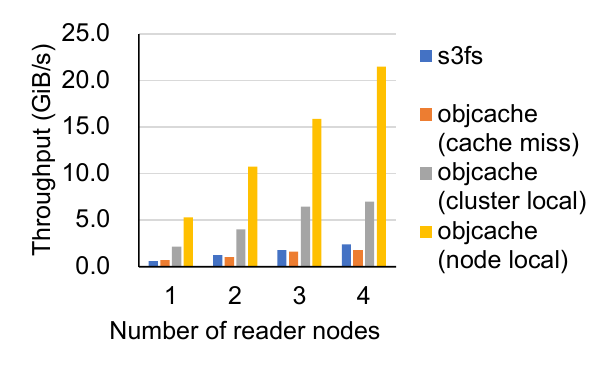}
  \caption{FIO sequential read throughput with cache misses and hits. We compared objcache performance with S3FS that wraps the same S3 bucket.}
  \label{fig: fio-s3fs}
\end{figure}

First, we show the baseline performance of objcache by generating synthetic file-intensive workloads with FIO.
In this experiment, we compared objcache to S3FS that wraps the same S3 bucket.
S3FS was configured to use the Linux page cache, 1-GB prefetch, 52-MB chunks, and 20 parallelisms for multipart uploads.
Objcache used 48-GB local NVMe, 16-MB chunks, 1-GB prefetching from external storage, and 256MB prefetching from cluster-local caches.

Figure~\ref{fig: fio-s3fs} shows the results of sequential reads when the cache misses, hits on cluster-local, or hits on node-local.
FIO used \verb|psync| I/O engine with a 4-GB file and the 8-KB block size, and 8 parallel jobs per node.
Objcache with cache misses showed up to a 27\% slowdown compared to S3FS since the detached deployment required overhead for networking.
However, cluster-local and node-local cache showed from 193\% to 1115\% higher than S3FS.
These results imply that utilizing node-local cache is a key to optimal performance of workloads.
As done in existing work, we may need a scheduler with an awareness of data locality or better prefetching techniques with profiling or other external information.

\subsection{Consistency and deployment models}
\label{sec: exp-fio}

\begin{figure}[t]
  \centering
  \begin{subfigure}{0.48\linewidth}
    \centering
    \includegraphics[width=\linewidth]{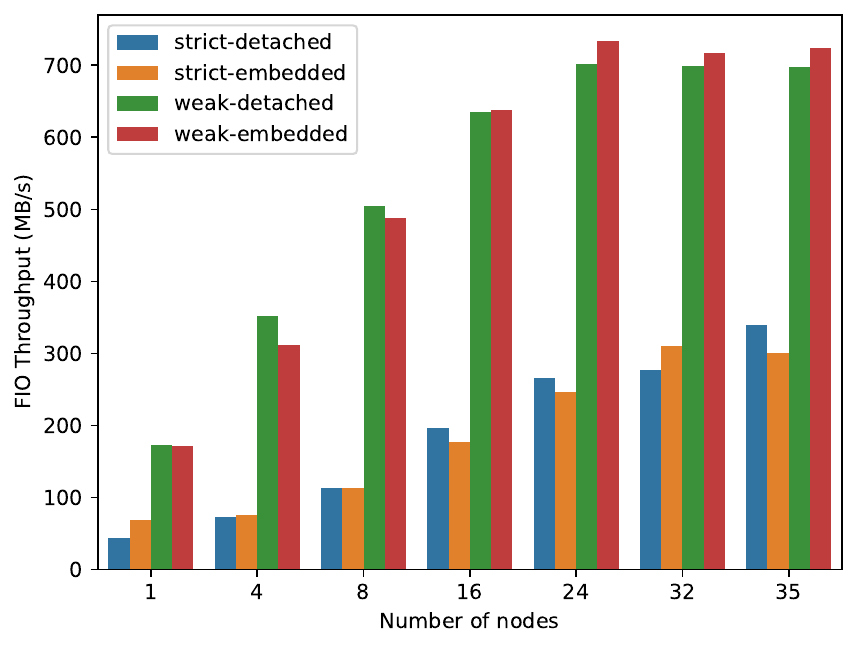}
    \caption{Sequential write throughput with different consistency and deployment models.}
    \label{fig: exp-fio-write}
  \end{subfigure}
  \hfill
  \begin{subfigure}{0.48\linewidth}
    \centering
    \includegraphics[width=\linewidth]{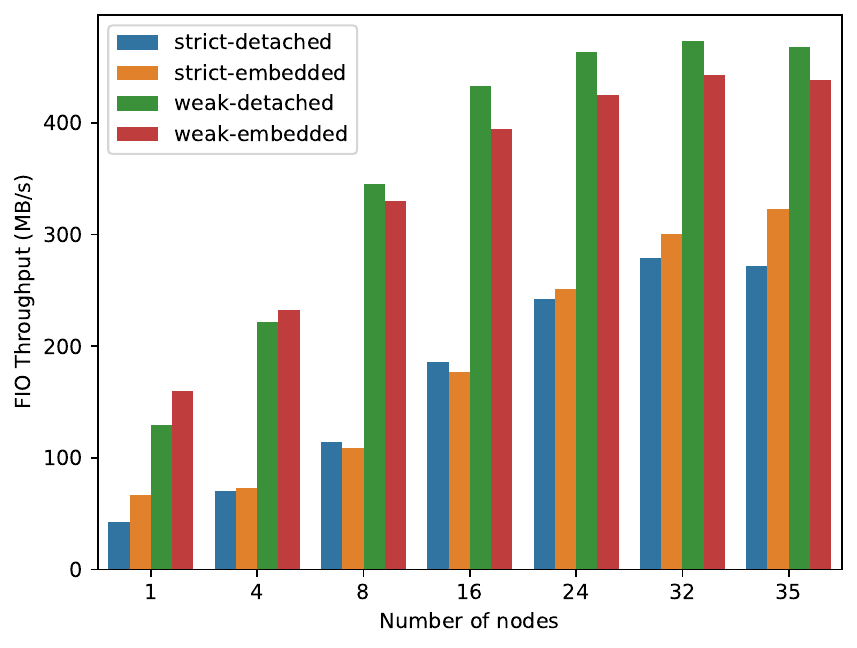}
    \caption{Random write throughput with different consistency and deployment models.}
    \label{fig: exp-fio-randwrite}
  \end{subfigure}

  \begin{subfigure}{0.48\linewidth}
    \centering
    \includegraphics[width=\linewidth]{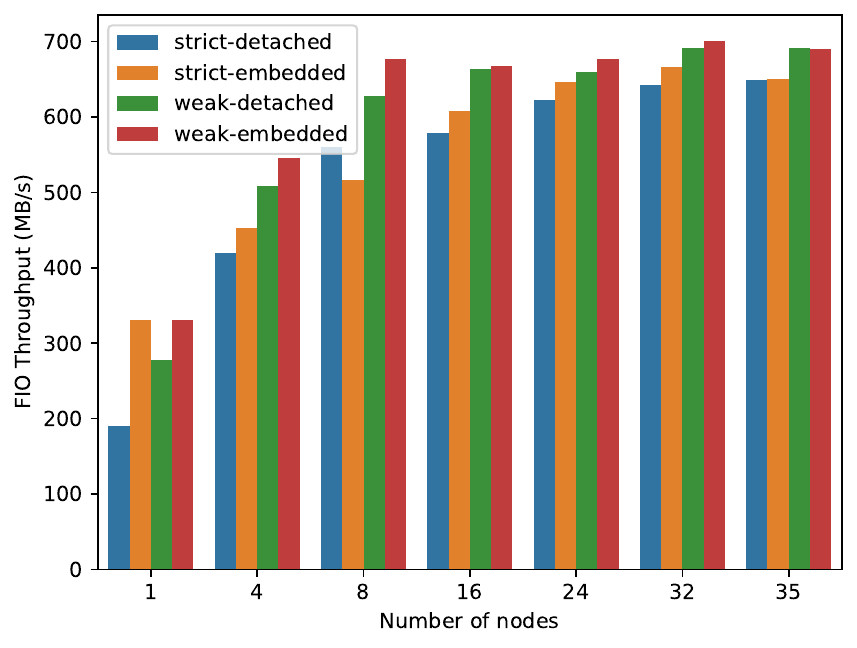}
    \caption{Sequential read throughput with different consistency and deployment models.}
    \label{fig: exp-fio-read}
  \end{subfigure}
  \hfill
  \begin{subfigure}{0.48\linewidth}
    \centering
    \includegraphics[width=\linewidth]{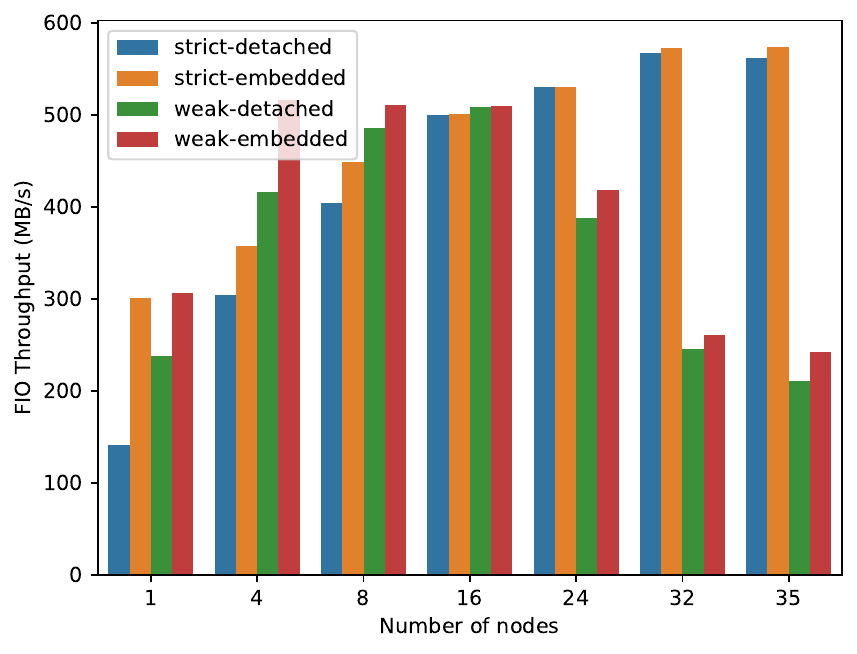}
    \caption{Random read throughput with different consistency and deployment models.}
    \label{fig: exp-fio-randread}
  \end{subfigure}

  \begin{subfigure}{0.48\linewidth}
    \centering
    \includegraphics[width=\linewidth]{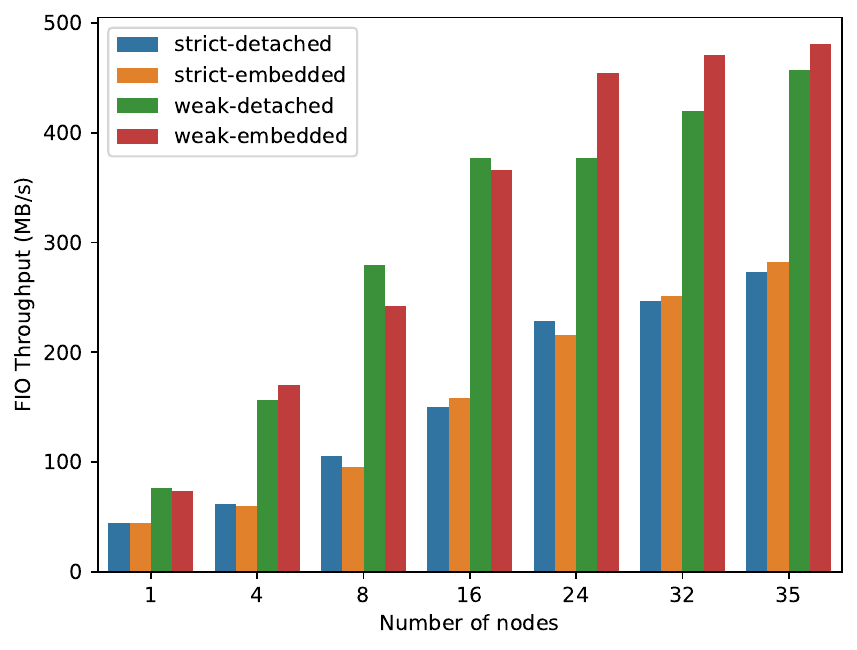}
    \caption{Throughput of sequential write with fsync().}
    \label{fig: exp-fio-write_fsync}
  \end{subfigure}

  \caption{Summary of FIO with different configurations. We tested the consistency of read-after-write (strict) and close-to-open (weak) with different deployment models (embedded and detached)}
  \label{fig: fio}
\end{figure}

In this section, we characterize the objcache performance of consistency and deployment models.
Our experiments run FIO with \verb|psync| I/O engine and 128KB block sizes to read/write a 1GB file per thread.
For experiments of the detached deployment model, we used a node in the 36-node cluster.
We set up different sizes of cache servers (from 1 to 35) and run the same number of threads as cache servers.
For sequential and random reads, we do not fill caches for FIO and so, all the results in this section show cache-miss performance.

Figure~\ref{fig: fio} shows performance characteristics of the consistency model and deployment models of objcache.
Write performance results (Figure~\ref{fig: exp-fio-write} and \ref{fig: exp-fio-randwrite}) showed that the strict consistency model degraded overall throughputs except for random reads.
Objcache changes the consistency model by turning on and off local buffering so that \verb|write()| is immediately reflected in cluster-local caches.
So, the weak consistency model improved write performance by buffering and batching multiple write requests.
We observed current Linux allowed up to 128 KB buffering even when we enabled the Linux page cache for the weak consistency model.
This implies that objcache (and probably other FUSE FSs) should show better performance on writes with the weak consistency model if Linux supports bigger buffering on FUSE.

The random read results (Figure~\ref{fig: exp-fio-randread}) showed an advantage of the strict consistency model in terms of performance, since it must simplify the buffering and caching at the client side.
Our prefetching mechanism was optimized for sequential reads, but it may incur overheads at non-uniform workloads.

Embedded deployment often showed better throughput since it can unify a FUSE and cache server in a node to bypass local networking between them.
However, random writes (Figure~\ref{fig: exp-fio-randwrite}) showed the opposite results at more than 8 nodes and the weak consistency model.
A possible reason is that the client and server shared buffers in memory at the weak consistency model, but it caused high memory pressure in the node.

\subsection{Start up of Model Serving}

We examine the startup latency of Triton~\cite{Triton}, a model serving infrastructure.
Triton can easily increase/decrease the number of servers according to its load.
With default settings, it loads models from target storage to GPU memory at its startup to get ready for serving requests.

We configured the target storage with a COS bucket and two wrapper FSs (S3FS and a single objcache server with the detached deployment).
Both S3FS and objcache used the same chunk size (16MB) with 64 parallel reads as prefetching (1GB in total).
S3FS enabled the Linux page cache but did not use local storage.
They stored a T5 model with 11 billion parameters~\cite{T5-JMLR20}.
We converted the original model file and generated 464 files with 43 GB in total for Triton's backend~\cite{Triton-FasterTransformer}.

\begin{figure}[t]
  \centering
  \includegraphics[width=0.7\linewidth]{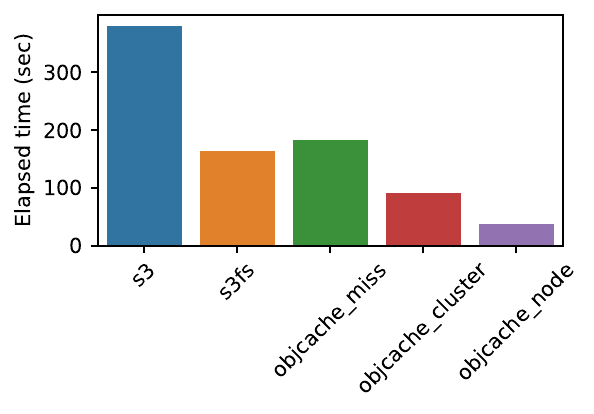}
  \caption{Average startup time of a model serving instance.
  Objcache varied its performance for cache misses (objcache\_miss),
  cluster-local cache hits (objcache\_cluster), and node-local cache hits (objcache\_node).
  We compared them with direct accesses of the S3 bucket (s3) and S3FS that wraps it (s3fs).}
  \label{fig: exp-modelserving-elapsed}
\end{figure}

Figure~\ref{fig: exp-modelserving-elapsed} shows that objcache speeds up the startup of Triton.
The time for download dominated the startup time of Triton.
Download time was critical for service availability since it was much larger (more than minutes) than an inference latency (less than 1 sec).
Triton can directly access S3 buckets, but it showed the worst performance (379.7 seconds) although we configured storing files at high-performance NVMe SSD.
Objcache improved the latency of the startup time when file requests were hit at cluster- or node-local cache (92.3 and 38.4 seconds, respectively).
S3FS showed a lower average latency (164.5 seconds) than objcache with cache misses (183.4 seconds), but it cannot share downloaded files among nodes.
So, if Triton scaled to multiple nodes, then S3FS required more time to pull files at every node than objcache with cluster-local caches.

The logic of GPU memory load was shared among different data sources including direct S3 accesses and S3FS/objcache.
So, the direct use of S3 had an overhead to copy all the files from an external bucket to local FS.
But also it could not efficiently use the CPU cache because copied files were evicted from the cache by the next file copy.
Wrapper FS like S3FS and objcache could reuse CPU cache during copies since they fetched files just before they were loaded to GPU memory.
In other words, the size of chunks should be configured to be less than the size of the L3 CPU cache to reduce copy overheads.
Note that our experiments for two clusters used CPUs with 32-MB L3 cache.
We also tested S3FS with 52-MB chunks with 20 parallel reads and it showed higher latency of startup time (218.9 seconds), although the first FIO experiment showed better throughput with the chunk size.

Another problem of direct S3 uses was that copied files were duplicated among cluster storage to consume disk capacity.
It could cause poor scalability to the disk consumption of models.
Objcache has the potential to scale to many large models if we add local storage and run new servers for it.

\subsection{Training Workload}

As an example use-case, we also tested a fine-tuning of T5 for automated grammar correction.
We compared objcache with the embedded deployment and S3FS for this workload to read pre-trained models and write checkpoint files.
We saved the T5 XXL model version 1.1 (42GB) and 34-GB Apache Arrow files to a COS bucket and wrapped it with the two FSs for the same training job.
Our goal is to test storage performance, so, we configured training to checkpoint every 32 iterations and stopped it at 128 iterations.
Every job container running at four nodes mounted S3FS or an objcache cluster running at each node.

\begin{figure}[t]
  \centering
  \includegraphics[width=\linewidth]{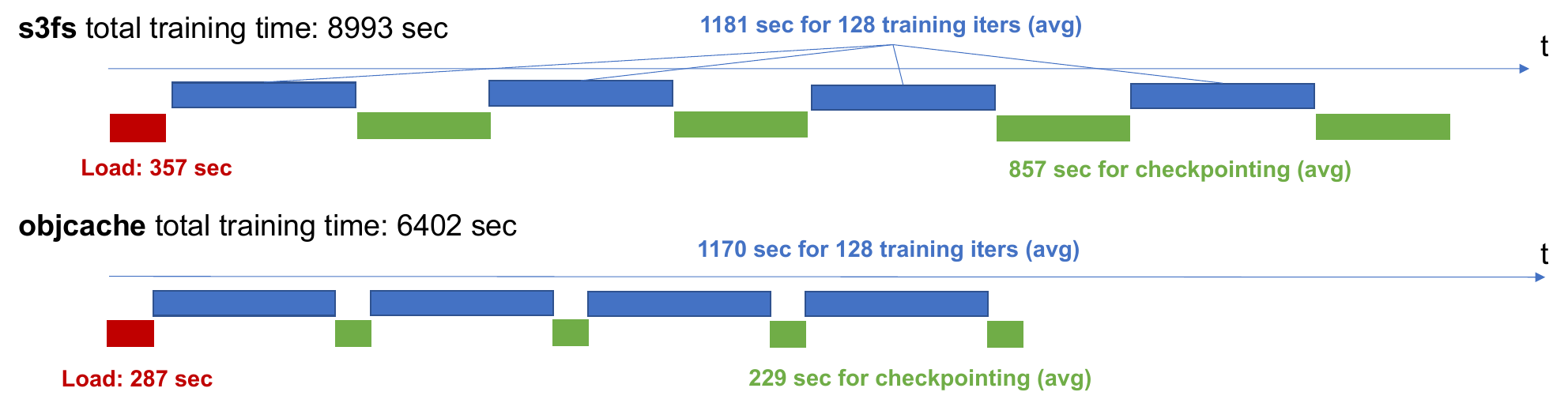}
  \caption{Breakdowns of training workloads around storage I/O. The red bar shows elapsed time for model loading. The green bar is elapsed time for checkpointing. The blue bar is the time for GPU computation.}
  \label{fig: exp-t5-gc}
\end{figure}

Figure~\ref{fig: exp-t5-gc} describes breakdowns of training workloads with S3FS and objcache.
Objcache reduced the time for loading the pre-trained model by 70 seconds (24\% speedup).
In this training workload, distributed jobs load the file in parallel, and thus, objcache can deduplicated downloads.
Objcache also reduced the time for checkpointing by 628 seconds (274\% speedup).
This is because S3FS synchronously uploaded files at every close, while objcache did it asynchronously.
As a result, file uploads overlapped with the GPU computation.

\subsection{Elasticity}
\label{sec: exp-elasticity}

We run experiments of starting up and shutting down a 36-node cluster.
Objcache should be configured to persist dirty files at some moderate periods, so it should not need large data migrations during cluster reconfiguration.
However, we examine pessimistic, synthetic scenarios where 1024 new files under 32 directories become dirty within a flush window before/after cluster reconfiguration.
In that case, objcache needs to migrate dirty metadata, chunks, and updated directories to a new predecessor node.
For node join experiments, we first run a single node cluster, write 1024 files, and add 35 nodes.
In contrast, node leave experiments start a 36-node cluster, write 1024 files, and remove a node one by one until all of them are deleted (i.e., scaling down to zero).
We used FIO random writes to generate the files from 1MB to 8MB sizes (4608MB in total).

\begin{figure}[t]
  \centering
  \begin{subfigure}{0.48\linewidth}
    \centering
    \includegraphics[width=\linewidth]{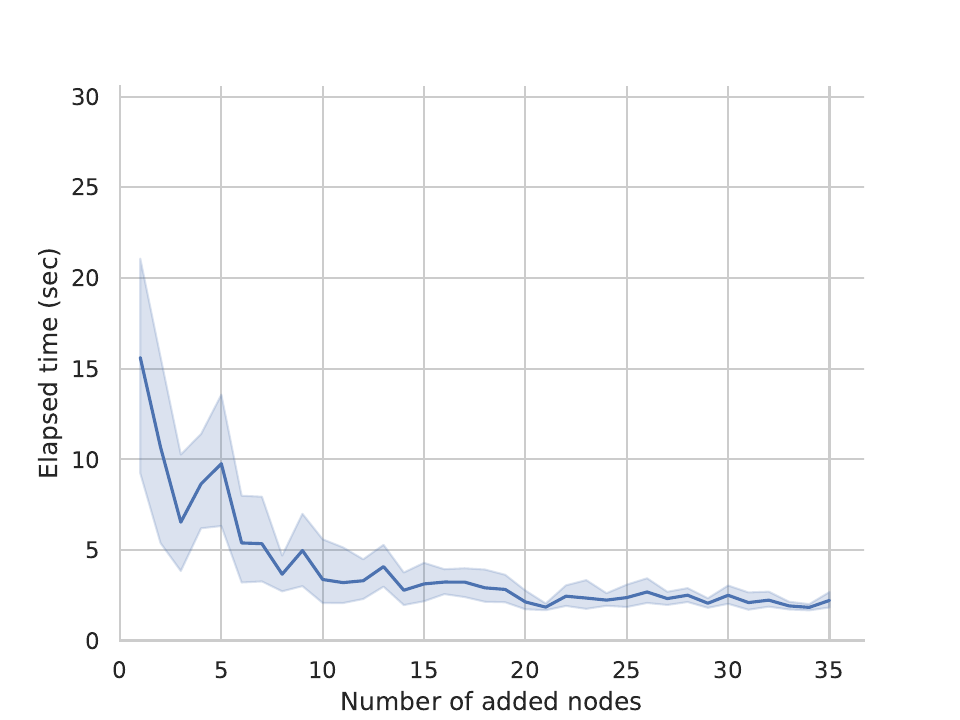}
    \caption{Time for scaling up from 1 to 36 nodes with dirty files.}
    \label{fig: exp-join}
  \end{subfigure}
  \hfill
  \begin{subfigure}{0.48\linewidth}
    \centering
    \includegraphics[width=\linewidth]{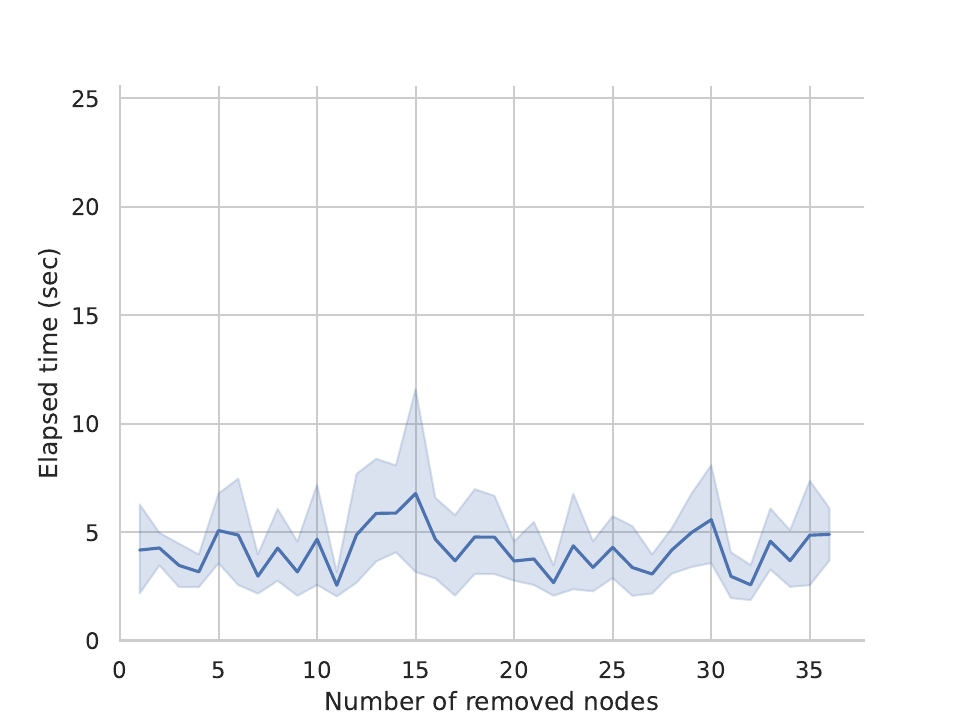}
    \caption{Time for scaling down from 36 to 0 nodes with dirty files.}
    \label{fig: exp-leave}
  \end{subfigure}

  \begin{subfigure}{0.48\linewidth}
    \centering
    \includegraphics[width=\linewidth]{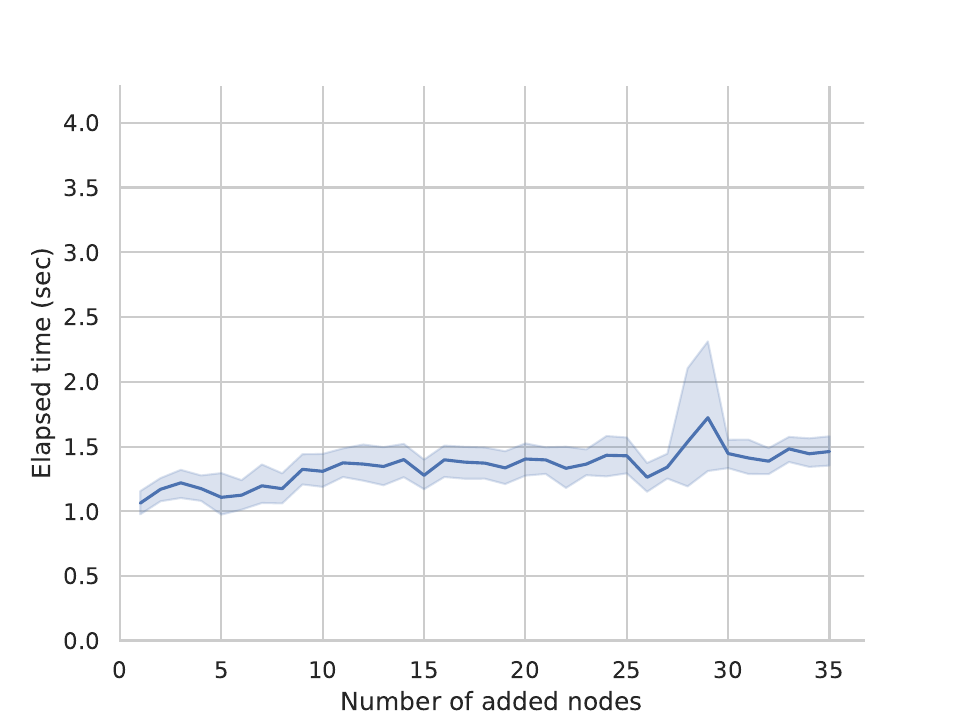}
    \caption{Time for scaling up from 1 to 36 nodes without dirty files.}
    \label{fig: exp-join-nodata}
  \end{subfigure}
  \hfill
  \begin{subfigure}{0.48\linewidth}
    \centering
    \includegraphics[width=\linewidth]{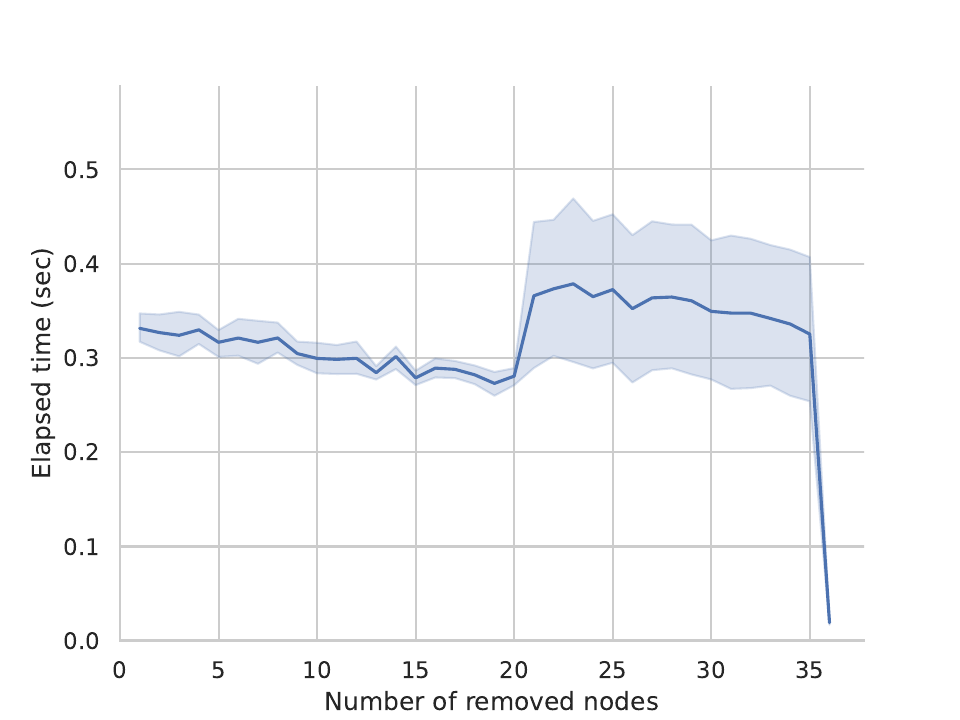}
    \caption{Time for scaling down from 36 to 0 nodes without dirty files.}
    \label{fig: exp-leave-nodata}
  \end{subfigure}
  \caption{Summary of scaling performance.}
  \label{fig: exp-joinleave-time}
\end{figure}

\begin{figure}
  \centering
  \begin{subfigure}{0.48\linewidth}
    \centering
    \includegraphics[width=\linewidth]{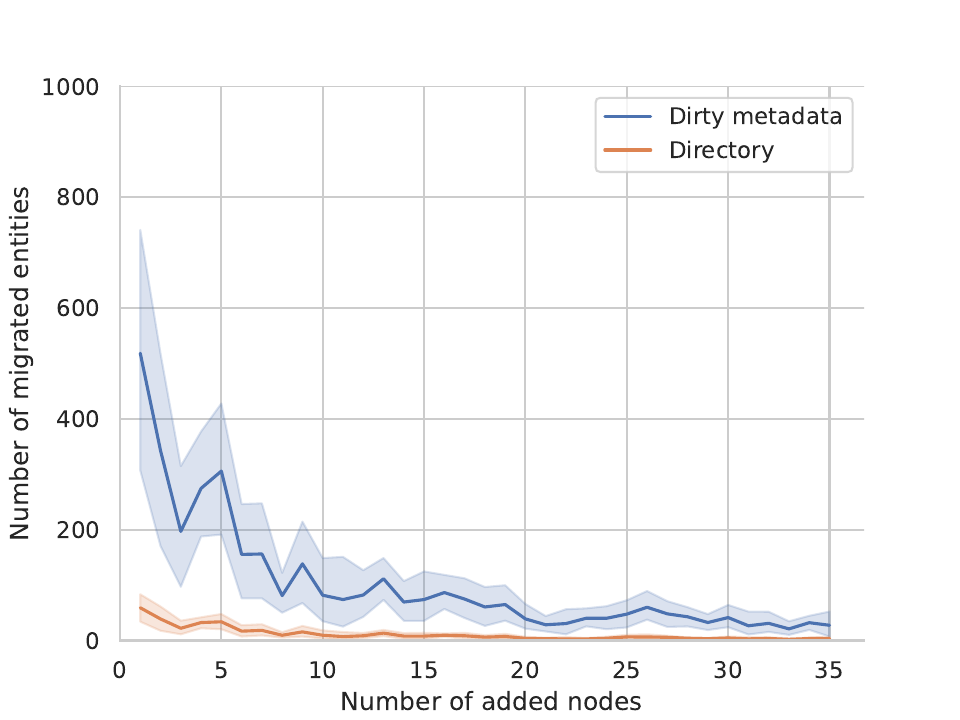}
    \caption{Number of migrated entities per node addition.}
    \label{fig: exp-join-details-meta}
  \end{subfigure}
  \hfill
  \begin{subfigure}{0.48\linewidth}
    \centering
    \includegraphics[width=\linewidth]{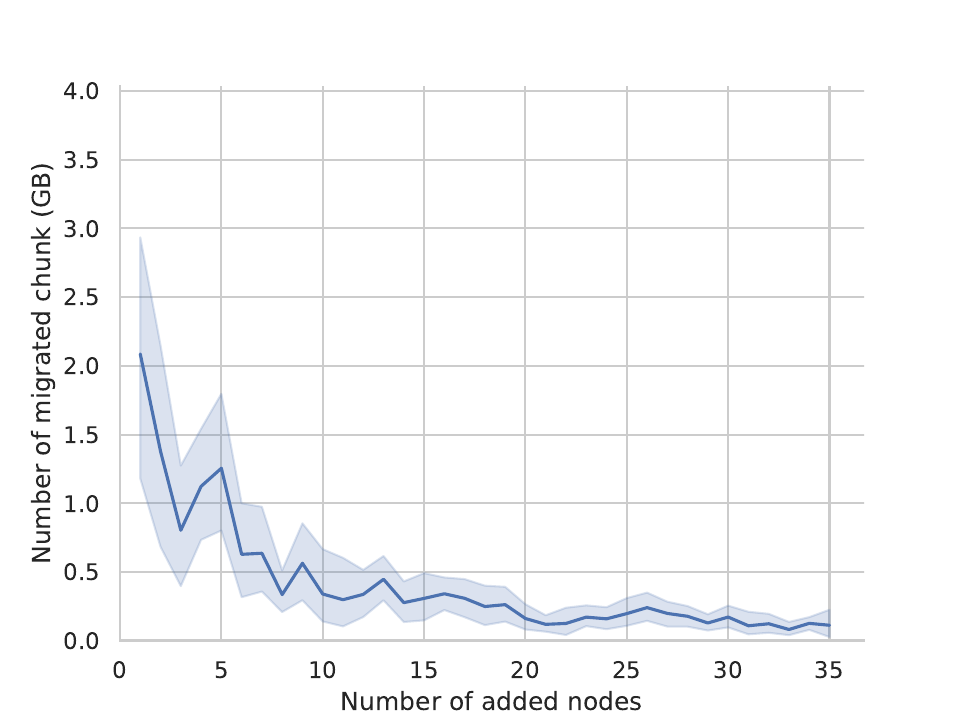}
    \caption{Total migrated chunk size per node addition.}
    \label{fig: exp-join-details-chunk}
  \end{subfigure}
  \caption{Details of migrated data.}
  \label{fig: exp-details}
\end{figure}

Figure~\ref{fig: exp-joinleave-time} summarizes average times for scaling objcache servers.
For comparison, it also shows a time for initializing a 36-node cluster without dirty files.
We can observe migration overheads by comparing two figures with or without dirty files (e.g., Figure~\ref{fig: exp-join} and \ref{fig: exp-join-nodata}).

Scaling up cause migration overheads especially when a cluster is small.
For example, Figure~\ref{fig: exp-join} shows that the first node addition required more than 15 seconds.
The first node addition migrated 518 dirty files, 59 directories, and 2-GB chunks (Figure~\ref{fig: exp-join-details-meta} and \ref{fig: exp-join-details-chunk}).
However, the cost of migration decreased in bigger clusters since the probability of a predecessor's change became relatively lower than in smaller clusters.
The 35-node cluster added a node within 3 seconds.

Figure~\ref{fig: exp-join-nodata} shows that a startup time of a cluster slightly increased when the cluster size increased.
This was because our transaction protocol synchronized the entire node list to every node, so the number of transaction participants was larger than in smaller clusters.
We believe that the synchronization overhead is negligible since the figure shows that adding a node required up to 2 seconds on average.
This experiment measures the time to add a node one by one, but users can start multiple nodes at a time to join them in parallel since the transaction ensures the ordering and consistency of the node list.
We also tested the creation time for 36 clusters in parallel, and the average time was less than 5 seconds.

Figure~\ref{fig: exp-leave} shows the performance of scaling down.
Current objcache uploads dirty metadata and chunks at a node shutdown and migrates metadata for directories to a new predecessor.
Compared with scaling up, scaling down showed relatively stable overheads at each node removal due to the random distribution of metadata and chunks.
Its overhead was more than 2 seconds but up to 6.8 seconds.
Figure~\ref{fig: exp-leave-nodata} shows that node removal took less than 1 second.
The removal of 36 nodes means zero scaling, which did not need a transaction and was completed in 19.2 milliseconds.

We believe current overheads of cluster reconfiguration can be acceptable for many use cases.
During migration, applications need to wait for reading and writing files.
However, we can still improve application availability by leveraging the pre-/post-copy migration techniques~\cite{LiveMigration-NSDI05}.

\section{Conclusion}
\label{sec: conclusion}

In this paper, we presented objcache, an elastic FS over external storage for container clusters.
A key technology of objcache was our transaction protocol to maintain the consistency of internal and external states.
Objcache offered flexible deployment and consistency models to meet different performance goals and administrative constraints.
Users can utilize objcache as a shared FS over COS, distributed storage tiering, and elastic storage in container clusters.

The main topic of this paper was consistency management with moderate performance, but we still have room for speedups as discussed in Section~\ref{sec: exp}.
Raft-based replication is also our future work.
We already designed objcache to enable replication, but need careful experiments and coding for realistic use cases that require high availability.

\printbibliography

\end{document}